\documentclass{PoS}

\title{Compact binary mergers: an astrophysical perspective}
\ShortTitle{Compact Binary Mergers}

\author{\speaker{Stephan Rosswog}\thanks{A footnote may follow.}\\
        School of Engineering and Science, Jacobs University Bremen,
        D-28759 Bremen\\
        E-mail: \email{s.rosswog@jacobs-university.de}}

\abstract{This paper reviews the current understanding of double neutron star and
neutron star black hole binaries. It addresses mainly (nuclear) astrophysics aspects 
of compact binary mergers and thus complements recent reviews that have emphasized 
the numerical relativity viewpoint.
In particular, the paper discusses different channels to release neutron-rich matter into 
the host galaxy, connections between compact binary mergers and short 
Gamma-ray bursts and accompanying electromagnetic signals.}

\FullConference{11th Symposium on Nuclei in the Cosmos, NIC XI\\
                July 19-23, 2010\\
                Heidelberg, Germany}

\def\msun{M$_{\odot}$}
\def\Msun{M$_{\odot}$ }
\def\bi{\begin{itemize}}
\def\i{\item}
\def\ei{\end{itemize}}
\def\be{\begin{equation}}
\def\ee{\end{equation}}
\def\ben{\begin{enumerate}}
\def\een{\end{enumerate}}
\def\bea{\begin{eqnarray}}
\def\eea{\end{eqnarray}}

\def\ben{\begin{enumerate}}
\def\een{\end{enumerate}}

\begin{document}

\section{Introduction}
In this paper I review the current understanding of the mergers of double neutron stars (DNS) and neutron star black 
hole systems (NSBH). I will collectively refer to them as ``compact binary'' mergers, and thus exclude systems with 
white dwarf components. Only binary systems that merge under the influence of gravitational 
wave emission will be discussed. Dynamical collisions as they occur in dense regions such as the 
cores of globular clusters will not be discussed here. Two excellent reviews 
have recently appeared \cite{faber09,duez10a} which put their emphasis on numerical relativity, 
I want to round up the picture by focusing on the (nuclear) astrophysics aspects of compact binary mergers.\\
To date 10 binary systems are known where at least the mass function and the periastron advance are
consistent with both stars being neutron stars \cite{lorimer08}. Five of these systems have small
enough orbital separations so that the constant leakage of orbital angular momentum due to gravitational wave
emission will cause a coalescence within a Hubble time. Incompletely understood physical mechanisms, 
poorly known parameters and hard to quantify selection effects make it difficult to estimate the 
rates at which such events occur. Rates derived from the observed systems roughly agree with 
those  from population synthesis models, about 40 to 700 Myr$^{-1}$ in a Milky Way equivalent galaxy 
\cite{kalogera04a,belczynski07}, but with an uncertainty of about an order of magnitude  
in each direction. Even less secure is the rate for NSBH systems, to date none
has been observed and population synthesis studies have predicted values from an order of magnitude 
more \cite{bethe98} to about two orders of magnitude less \cite{belczynski07} than the DNS merger rate.\\
\noindent It is difficult to overrate the importance of this type of binary system:
\bi
\vspace*{-0.3cm}   
\i The first discovered system, PSR 1913+16, has delivered --via the measured decay of the binary   
    orbit-- the first unambiguous evidence for the existence of gravitational waves, in excellent 
    agreement with the predictions of  General Relativity \cite{taylor89}.
\vspace*{-0.3cm}   
\i The measurement of at least two Post-Keplerian parameters allows the measurements of 
   {\em individual} neutron star masses. For example, the masses of PSR 1913+16 are 
   $m_1= 1.4414$ \Msun and $m_2= 1.3867$ \Msun $\pm \; 0.0002$ \Msun \cite{weisberg05}.
    Accurately known masses provide stringent tests for the hadronic physics inside a neutron star
    \cite{lattimer04}. This regime of high densities, but low temperatures
    is hardly accessible to any laboratory experiment, but it can be probed via accurately known 
    neutron star masses.
\vspace*{-0.3cm}
\i  The large "compactness" of neutron stars, $\zeta \equiv G M_{\rm ns}/R_{\rm ns} c^2\approx 0.2$
   (for comparison: the compactness of the Sun is $\sim 10^{-6}$), and their high orbital velocities,
   $v \sim 10^{-3} c$ ($c$ being the speed of light) make DNS excellent laboratories for strong 
   gravity. They allow for accurate tests that have the potential to distinguish General Relativity from other 
   theories of gravity \cite{kramer09}.
\vspace*{-0.3cm}
\i The last stages of the inspiral of a DNS system are a prime candidate for the first {\em direct} detection
   by the ground-based gravitational wave detectors \cite{sigg08,arcenese08,grote08} that have now finished 
   their first complete science runs. Population synthesis models \cite{belczynski07} have predicted detection 
   rates for the Advanced LIGO project near 10 for DNS and around one event per year for NSBH systems. The
   detection efficiency could be further enhanced by the simultaneous detection of accompanying signals
   in other channels. 
\vspace*{-0.3cm}
\i The astrophysical production site of the r-process is still an unsolved problem. For many years supernovae, in particular
   the neutrino-driven winds from a new-born neutron star, were considered very promising sites to forge r-process material,
   but recent studies find it difficult to reproduce the observed abundance patterns with parameters that are considered
   plausible for core-collapse supernovae \cite{arnould07,roberts10}. The main competitor are the neutron-rich ejecta that 
   seem unavoidable in a compact binary merger \cite{lattimer76,lattimer77,rosswog99,ruffert01,oechslin07a,metzger08}.
\vspace*{-0.3cm}
\i Since the very beginning, compact binary mergers have been considered a prime candidate for the central engine
   of Gamma-ray bursts (GRBs) \cite{blinnikov84,paczynski86,goodman86,eichler89,paczynski91,narayan92}
   and they have survived being confronted with a wealth of observational results. Although the case is far from
   being settled, they still are the major candidate for the central engine of short GRBs 
   \cite{piran05,nakar07,lee07,gehrels09}.
\ei

\section{What are the challenges?}
Compact binary mergers are prime examples of multi-physics and multi-scale problems.
%gravity
With compactness values $\zeta$ of $\sim 0.2$ for neutron stars and 0.5 at the event
horizon of a Schwarzschild black hole, relativistic gravity has obviously a major impact on the dynamics of
close, compact binary systems.
During the final inspiral stages and the merger the space-time geometry changes on time scales that become comparable
to the compact object dynamical time scales of order 1 ms.\\
%strong interactions
Their equation of state determines how the neutron stars react on such space-time changes. Tidal heating
has no major impact on the bulk matter evolution \cite{lai94c} since thermal energies are tiny in comparison
to the relevant Fermi energies. In the merger remnant, however, temperatures can reach $\sim 10$ MeV and, 
dependent on the local
density, this may not be negligible at all. The neutron stars start out macroscopically from hydrostatic and,
microscopically, from cold $\beta$-equilibrium. The disruption decompresses a good fraction of the 
star across the phase transition to inhomogeneous nuclear matter where free nucleons and possibly nuclei and 
nuclear clusters coexist. Some of the debris is ejected, but most of it forms an accretion torus around 
the central object. Nuclear reactions during this and subsequent stages can have a number of observable 
consequences, see Sect.~\ref{sec:nu_wind}, \ref{sec:disk_disinteg} and \ref{sec:transients}.\\
In the merger remnant, the neutron to proton ratio evolution is regulated by the competition between 
electron and positron captures on the one side, and neutrino and anti-neutrino absorptions on the other. 
At the prevalent densities,
photons are completely trapped and neutrinos are the only viable cooling agents. Being released at tremendous 
luminosities, neutrinos can also be re-absorbed and have the potential to drive strong thermal winds which, in turn, may 
have a major impact on both nucleosynthesis, Sec. \ref{sec:mass_loss}, and the ability of compact binary mergers 
to possibly produce Gamma-ray bursts, Sec.\ref{sec:GRB}. \\
% magnetic fields
Neutron stars are naturally endowed with strong magnetic fields and a compact binary merger offers a plethora of possibilities 
to amplify them. They may be decisive for the fundamental mechanism to produce a GRB in the first place, but they may
also determine --via transport of angular momentum-- when the central object produced in a DNS merger collapses into a 
black hole or how accretion disks evolve further under the transport mediated via 
the magneto-rotational instability (MRI) \cite{balbus98}.\\
% the numerical challenge(s)
A numerical simulation of a compact binary merger is further complicated by several additional 
challenges. For example, the speed of sound, $c_{\rm s}$, inside a neutron star can easily reach tens of percent 
of the speed of light. For explicit schemes this severely restricts numerical time steps
via the Courant-Friedrichs-Lewy stability condition \cite{press92}, 
$\Delta t < 10^{-5} (\Delta x/1 {\rm km}) \; (0.3 c/c_s)$. This can become a major 
stumbling block for phenomena that substantially exceed the dynamical time scales of the central object. 
Another complication (that this field shares with other astrophysical fluid simulations) are sub-resolution
length scales on which physical instabilities such as Kelvin-Helmholtz or the MRI are seeded. 
In nature they would  feed back and leave a noticeable imprint on the large-scale 
dynamics. Ideally, such cases should be, but rarely are, treated via ``subgrid-models''. Further 
challenges can arise from numerical non-conservation of nature's conservation laws. For example, the 
orbital dynamics in a binary system is very sensitive to the transfer of angular momentum, both via tides 
and transferred mass. In unfortunate circumstances, purely
numerical effects like an artificial loss of angular momentum could swamp true physical effects of the same magnitude, say due
to the emission of gravitational waves. Another subtlety, mainly for grid-based codes, is that ``vacuum'' is difficult
to treat as such and it is generally mimicked by some low-density atmosphere. Although the corresponding 
background densities are usually many orders of magnitude lower than the peak (neutron star) densities, they 
can still easily exceed the densities inside a white dwarf and may thus have a major impact on questions that 
are related to low density regions, such as the dynamical ejection of matter or emerging winds. Such effects 
may also in part be responsible for the current relativistic NSBH calculations not yet converging with respect 
to resulting disk masses, see below.

% intrinsically multi-scale and multi-physics
\section{Current approaches}
\label{sec:approaches}
The last decade has seem a tremendous leap forward on essentially all fronts of the compact binary merger problem. 
After the first approaches with Newtonian gravity and polytropic equations of state (EOS) 
\cite{oohara89,rasio94,zhuge94,lee99a,lee99b,lee00} 
a bifurcation took place with one line of research focusing on the strong gravity aspects while neglecting non-gravity physics 
and another line addressing exactly the latter, but in an essentially Newtonian framework. Both lines obtained
substantial progress in the last decade, now the efforts focus on catching up with so far neglected physics aspects.\\
%\subsection{Gravity}
First steps beyond purely Newtonian gravity are the application of pseudo-relativistic potentials \cite{paczynski80}
to NSBH systems \cite{lee99b,rosswog05a} and the development of  post-Newtonian (0+1+2.5 PN) hydrodynamics codes
\cite{ayal01,faber00}. The latter approaches were somewhat hampered by the limitations of the 1PN approach in a 
neutron star context, since higher order PN corrections are not necessarily small.
The conformal flatness approximation (CFA) \cite{isenberg08,wilson95,oechslin02,faber06a} represents a further step towards 
solving the full GR equations. It assumes that the spatial part of the metric, $g_{ij}$, is, and remains, conformally flat: 
$g_{ij}= \Psi^4 \delta_{ij}$, where $\Psi$ is the conformal factor. To evolve the space-time dynamically, one needs to solve
five coupled, non-linear partial differential equations with non-compact source terms. The CFA is substantially faster 
than full GR (but substantially slower than Newtonian gravity) and therefore, for the same available computing resources, 
allows to invest more time in hydrodynamic resolution or other physics ingredients of the problem. CFA is exact for 
spherical symmetry and at least accurate to 1PN order in the general case. For spinning neutron stars, it has been 
shown to be accurate to a few percent \cite{cook96}, for general situations the accuracy is not well known and difficult 
to determine. While overall being a very useful approach, the CFA is restricted by the assumption of a waveless 
space-time which, at least in principle, leads to formal inconsistencies if gravitational wave backreaction terms 
are applied to drive the binary towards inspiral and merger. \\
The first fully general relativistic merger calculations were performed by Shibata and collaborators 
\cite{shibata00,shibata02,shibata03}. Today, two GR formulations are commonly used, the so-called BSSN-formulation 
\cite{shibata95,baumgarte99} and the generalized harmonic formalism
\cite{friedrich85,garfinkle02,bona03,pretorius05,lindbloom06}. In both cases, certain first order derivatives are promoted 
to the status of independent functions and this allows for the numerically stable evolution of dynamical space-times. To date, many
numerically stable evolutions of compact binary systems have been carried out in full GR, e.g. 
\cite{baiotti05,shibata07,baiotti08,duez08}. In hybrid approaches with a fixed Kerr space-time and Newtonian 
self-gravity \cite{rantsiou08} black hole spin effects on the NSBH merger dynamics were studied. These effects
have recently also been addressed in full GR \cite{foucart10}.\\
%\subsection{Other physical processes}
% EOS
But as outlined above, compact binary mergers are not only a dynamical strong gravity problem.
They are, for example, heavily influenced by the nuclear EOS. Its stiffness varies substantially over 
the density range that is relevant for compact binary mergers. The effective adiabatic exponents 
range from beyond 3 near nuclear densities to close to 4/3 for decompressed nuclear matter (for 
an illustration see, e.g. Fig. 5 in \cite{rosswog02a}) and this leaves a pronounced imprint on 
the matter evolution (compare, for example, Figs. 1 and 2 in \cite{rosswog99}). Temperature and 
composition dependent nuclear equations of state, mainly those of Lattimer and Swesty
\cite{lattimer91} and Shen et al. \cite{shen98a,shen98b}, have been heavily used in a variety of studies 
\cite{ruffert96,ruffert97a,rosswog99,rosswog02a,price06,oechslin07a,duez10b}. Some studies have also
explored the impact of strange matter on the merger outcome \cite{lee01b,bauswein09,bauswein10}.\\
% neutrinos: leakage, transport
For grid-based hydrodynamics, opacity-dependent neutrino leakage schemes in compact binary merger simulations have 
been pioneered by \cite{ruffert96}. In \cite{rosswog03a} a leakage scheme was suggested that does not make use
of average neutrino energies, but instead determines the neutrino emission by integrating over a neutrino
energy distribution. This scheme  was applied in SPH merger simulations \cite{rosswog03c,price06} and has
recently been generalized for the use in general relativistic calculations \cite{sekiguchi10b}. While these
leakage schemes have turned out to be encouragingly accurate in describing the $\nu$-cooling of matter 
\cite{dessart09}, they do not account for heating effects from neutrinos that were emitted at a different location. 
Therefore, they do not allow to study phenomena such as neutrino-driven winds. The latter have been explored
recently with more elaborate neutrino transport methods \cite{dessart09}.
% disk studies
Effects from the equation of state and neutrino cooling on remnant disks have also been
intensively studied \cite{lee02,lee04,lee05b}.\\
% magnetic fields
The past few years have further seen the exploration of the magnetic field evolution in compact binary 
mergers, both in the Newtonian \cite{price06,rosswog07c} and the general-relativistic case 
\cite{shibata06a,anderson08b,liu08,giacomazzo10}.

% nuclear reactions

\section{The emerging patchwork picture}
As outlined above, the merger phenomenon is physically far too complex for any of today's
models to address all interesting aspects accurately at the same time. We are thus left with a 
"patchwork picture": for each question we have to pick  the best available approach 
in the hope  that the imperfect aspects do not substantially alter the 
conclusions.

\subsection{Mass loss and nucleosynthesis}
\label{sec:mass_loss}
The enrichment of the Galaxy with neutron-rich material during a compact binary merger
is not only possible, but, on the contrary, rather difficult to avoid. Apart from the 
matter that is ejected dynamically by gravitational torques, there is an additional contribution 
due to neutrino-driven winds and last, but not least, neutron-rich matter that becomes dispersed
into space in the course of the viscous evolution of a neutron star debris disk.
While the initial starting point is the same, cold neutron star matter in $\beta$-equilibrium,
the three channels differ in the amounts of released matter, in their entropies, expansion 
time scales and electron fractions. Therefore they likely produce a different nucleosynthetic signature.

\subsubsection{Dynamic ejection}
\label{sec:dyn_ej}
\noindent In their study of neutron star black hole binaries Lattimer and Schramm \cite{lattimer74} had 
found that the "Roche limit", where the neutron star's
self-gravity has to surrender to the tidal forces exerted by the black hole, lies outside the event 
horizon. They estimated that a fraction of about $\sim 0.05$ of the neutron star could become unbound, 
and realized that, folded with the estimated neutron star black hole merger rate, the amount of ejected
material would be comparable to the estimated r-process inventory of the Galaxy. They concluded
that the most important observational consequence of a neutron star black hole 
encounter may be nucleosynthesis \cite{lattimer76} and further speculated that also double neutron stars may enrich 
the Galaxy with neutron-rich material. This latter topic has been discussed in more detail together with the production
mechanism of GRBs  in \cite{eichler89}.\\
Early Newtonian hydrodynamic simulations \cite{rosswog99} that made use of a nuclear equation of 
state \cite{lattimer91} found dynamically ejected masses between $4 \times 10^{-3}$ and $4 \times 10^{-2}$ 
\msun, depending on the initial neutron star  spins. They also noted a strong sensitivity to the stiffness of 
the nuclear EOS \cite{rosswog00} with too soft an EOS leading to no resolvable mass loss. Recent simulations 
\cite{oechslin07a} that make use of the conformal flatness approximation and explore several nuclear equations 
of state  find an ejecta range from $\sim 10^{-3}$ up to a few times $10^{-2}$ \msun, with the 
arguably most common case (1.38 and 1.42 \Msun with negligible neutron star spins) ejecting about $3 \times 10^{-3}$ 
\msun. More asymmetric systems show the tendency to eject larger amounts of material, therefore, 
in nature double neutron star mergers should produce a distribution of dynamical ejecta, essentially set by 
the binary mass distribution. With the estimated rates \cite{kalogera04a}, the 
dynamical ejection alone could enrich the Galaxy by an amount of matter that is 
comparable to its estimated r-process inventory, $\sim 10^5$ \msun.

\subsubsection{Neutrino-driven winds from merger remnants}
\label{sec:nu_wind}
\noindent Similar to the proto-neutron star case \cite{duncan86}, 
the huge neutrino luminosities  from the remnant of a compact object merger 
ablate a substantial amount of baryons from its surface. It had been realized early on that such a 
neutrino-driven wind holds much promise as a possible nucleosynthesis site \cite{eichler89,ruffert97a,rosswog03c}, 
but also that it could pose a serious threat to the emergence of the ultra-relativistic outflow that is required 
to power a GRB.\\
A compact binary merger results \cite{ruffert97a,rosswog99,ruffert01,rosswog02a} in a central remnant consisting of  
either a black hole or, if the initial ADM mass of the binary was smaller than a threshold mass of 1.35 times 
the maximum mass of a cold, non-rotating neutron star \cite{shibata06c}, it could produce a meta-stable, neutron 
star-like central object. The recent determination of the neutron star mass in J1614-2230 with 
$M_{\rm ns}= 1.97 \pm 0.04$ \Msun \cite{demorest10} places this threshold mass to at least 2.66 \Msun and 
therefore many, maybe most (depending on the binary mass distribution), merging DNS could pass through a phase 
with a metastable central object. This central object is surrounded by a disk of $\sim 0.1$ \msun, the detailed 
value depending on spin, mass ratio and nuclear EOS, composed of mainly free nucleons at a temperature 
of a few MeV. Some fraction of material (0.02 - 0.08 \msun)  \cite{rosswog07a} is sent into nearly unbound 
orbits that will fall back at a later time and a smaller mass fraction, see Sec.~\ref{sec:dyn_ej}, that is 
dynamically ejected.\\
Nucleons in the inner parts of the disk at a distance $r$ from the center are gravitationally bound with an energy 
of $E_{\rm grav} \approx - 35 \; \rm{MeV} \; (M_{\rm co}/2.5$ \msun) $(100$ km/$r$), to the central object of mass 
$M_{\rm co}$, i.e. $|E_{\rm grav}|$ in the remnant is comparable to the typical 
neutrino energies \cite{ruffert97a,rosswog03a}. 
Therefore, chances are good that neutrinos can lift nucleons out of the remnant gravitational potential and drive
a wind. Such winds have been investigated as possible nucleosynthesis sites \cite{surman08,kizivat10}
in parametric studies, but a quantitative investigation of their evolution and geometry has only become feasible 
recently \cite{dessart09}. In the latter work the authors started from the 3D remnant structures as calculated in 
\cite{price06,rosswog07c} and further evolved them with a 2D neutrino-hydrodynamics code \cite{burrows07a}. 
As a byproduct, they scrutinized existing neutrino leakages schemes \cite{ruffert97a,rosswog03a} against more sophisticated 
transport methods and found them reassuringly accurate. The main result of the study was that charge-current neutrino 
reactions thermally drive a bipolar wind with a mass loss rate of $\langle \dot{M} \rangle \approx
10^{-3}$ \msun s$^{-1}$. In each merger neutrino-driven winds blow $\sim 10^{-4}$ \Msun of material with 
$Y_e \approx 0.1 - 0.2$ into the host galaxy (assuming a central object lifetime of $\sim 100$ ms). 
Magneto-rotational effects were not accounted for in this study, but are likely to substantially increase this mass loss
\cite{metzger07}. 
The above $\nu$-wind study assumed that the central neutron star remains stable during the time scale of the simulation. Since a 
substantial part of the mass loss stems from the surface regions of the central object (see Fig. 10  in \cite{dessart09}), 
an early collapse to a black 
hole could seriously influence the amount of mass loss by shutting off the central object contribution. This could be favorable 
for GRBs in allowing ultra-relativistic outflow to develop along the binary rotation axis. How the wind evolves after the formation
of a black hole  still remains to be explored in future studies.

\subsubsection{Disk disintegration}
\label{sec:disk_disinteg}
\begin{figure}[tbp] %  figure placement: here, top, bottom, or page
   \centering
   \centerline{
   \includegraphics[height=9.4cm,angle=-90]{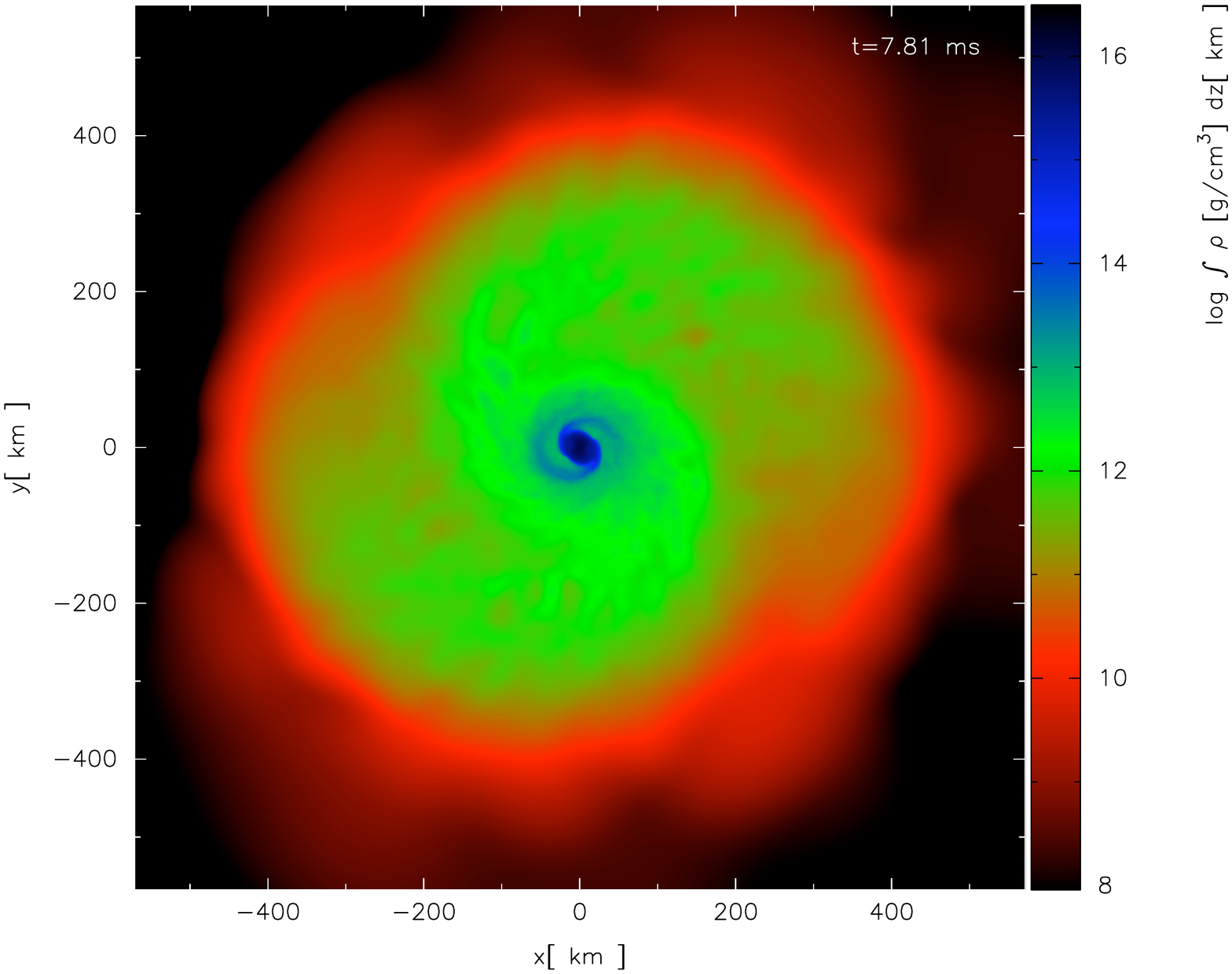} \hspace*{-1.2cm}
   \includegraphics[height=9.7cm,angle=-90]{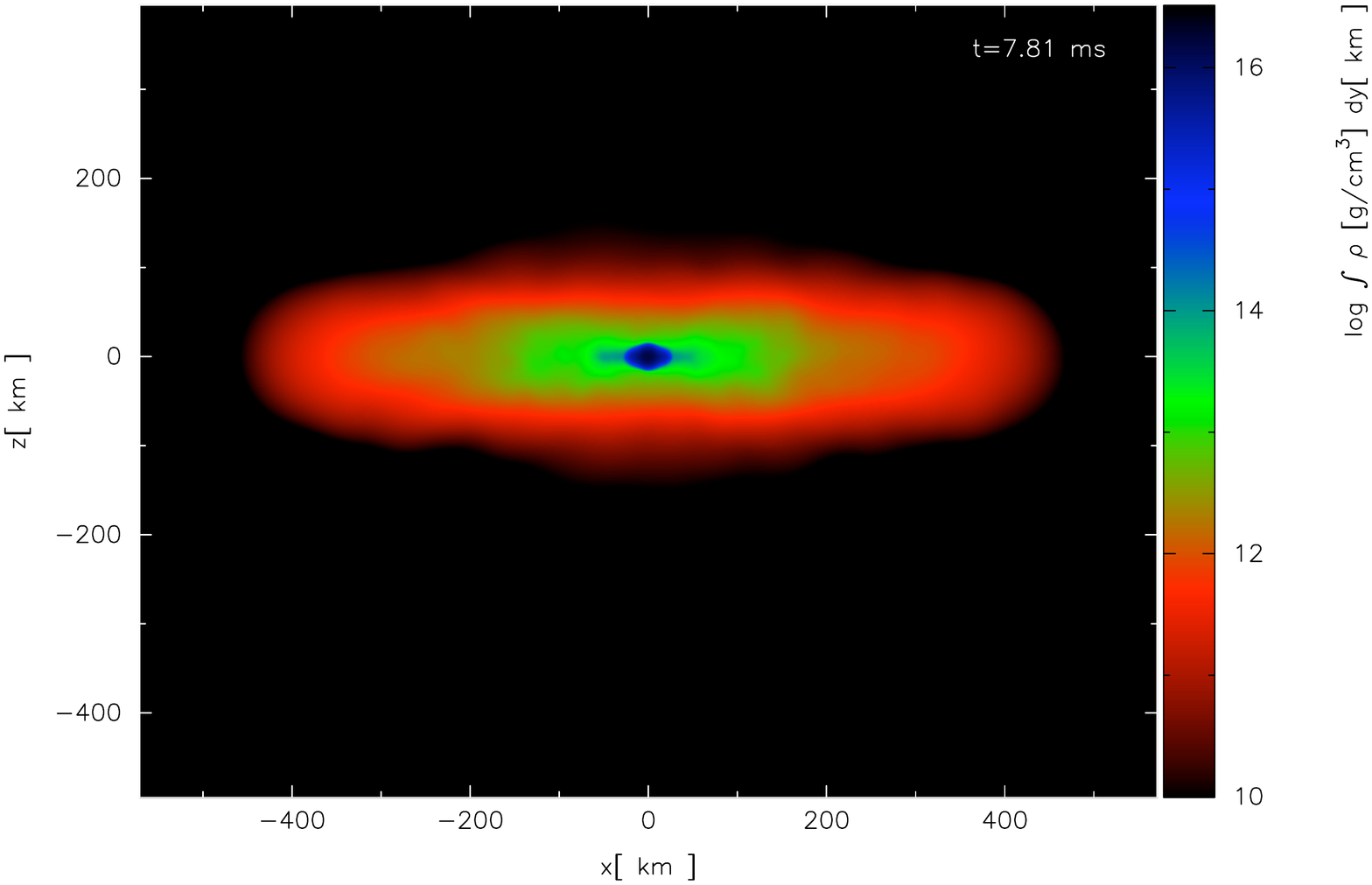} }
   \caption{Mass distribution resulting from an irrotational double neutron star system with
            with 1.4 \Msun per star. Shown is the column density distribution in XY- (left) 
            and XZ-plane (right) at about 5 ms after merger (7.81 ms after simulation start).}
   \label{fig:disk}
\end{figure}
\noindent Also the remnant disk of a compact binary merger can release neutron-rich matter into the galaxy, either
via viscous heating in an advective, thick disk \cite{metzger08} or via the energy release from the recombination 
of nucleons into nuclei \cite{chen07}.\\ 
The geometry of a post-merger configuration (2 x 1.4 \msun, no initial stellar spins) is shown in Fig.~\ref{fig:disk}. 
At this stage, the disk is not yet in an equilibrium state, where
radial velocities are entirely determined by the viscous transport of angular momentum. In our (Newtonian) simulations
the disk extends after a few milliseconds out to $\sim$  500 km (somewhat dependent on EOS, the initial stellar 
spins and the mass ratio) and possesses temperatures of $\sim$ 3 MeV  at the inner edge. GR effects tend to 
decrease disk mass and size and to increase the temperatures. The decompression occurs too rapidly for 
the weak interactions to instantly reach $\beta$-equilibrium and therefore the initial disk possesses the (advected) $Y_e$ 
values from the original neutron star, typically close to 0.05. After the initial transient stage, the disk 
evolution is driven by
the viscous coupling between annuli of neighboring radii which transports angular momentum outwards and allows
matter to be accreted onto the central object. At the same time, the
disc spreads on a viscous time scale $\tau_{\rm visc} \approx \frac{1}{\alpha \Omega} (r/H)^2$. Here $\alpha$
is the Shakura-Sunyaev viscosity parameterization \cite{shakura73}, $\Omega$ the angular frequency and $H$ the disk scale height.
If a central, hypermassive neutron star is present, matter piles up rapidly around it and forms 
a hot bulge with neutrino optical depths $\tau_{\nu}$ of a few \cite{rosswog03a}. For non-spinning black holes, 
the densities and optical depths are lower, but they can reach the previous values for spin parameters
$a$ in excess of 0.9 \cite{chen07}. Apart from possibly very close to the center, the inner disk regions are efficiently 
neutrino-cooled and therefore geometrically thin, $H/r\sim 0.2$. As the disk spreads further and becomes cooler, neutrinos
are no longer able to efficiently discharge the viscously released heat and the disk becomes geometrically thick
and advective \cite{beloborodov08,metzger09b}. Such advective disks are only marginally bound and expected to loose a
substantial fraction of their mass in a disk wind \cite{blandford99b}. 1D models \cite{metzger09b} indicate
that the disk masses evolve according to $M_{\rm d} \propto t^{-1/3}$ and that about 70\% of the initial disk mass has 
been accreted by the time the disk becomes advective. Thus, about 0.03 \Msun $(M_{\rm d,0}/0.1$\msun) of debris
material become unbound in a DNS merger. Fairly independent of the detailed viscous spreading, the released material is 
moderately neutron rich, $\sim 0.1 < Y_e < 0.5$ with a peak at $Y_e\approx 0.3$ and possesses entropies between 3 and 20 
$k_{\rm B}$ per baryon \cite{metzger09b}.\\
% now the burning+ release issues
This mass loss is most likely further enhanced by burning processes \cite{lee07,beloborodov08,lee09}. Outside of
$R_{\rm He}= 450 \; {\rm km} \; (M_{\rm co}/2.5$ \msun) the formation of alpha particles from free nucleons releases enough nuclear
energy to gravitationally unbind the disk. Thus a substantial part, if not most, of the late-time disk mass becomes 
dispersed into space. Such a burning-driven wind has also been suspected \cite{lee09} to suppress the mass supply to the 
central engine which could explain the shutting off of post-burst GRB activity that has been observed in some GRBs 
\cite{gehrels09}.

\subsubsection{R-process enrichment of the Galaxy}
\noindent As outlined above, a neutron star coalescence unbinds neutron-rich material in at least three different ways: 
\bi
\vspace*{-0.3cm}
   \i [i)] dynamic ejecta, $m_{\rm ej} \approx 3 \times 10^{-3}$ \Msun \cite{oechslin07a}, with entropies of 
           $s \sim 4 \; k_{\rm B}$ baryon$^{-1}$\cite{rosswog99} and $0.01 < Y_e < 0.5$ with most of the mass 
           possessing $Y_e \approx 0.05$ \cite{rosswog05a}, 
\vspace*{-0.3cm}
   \i [ii)] neutrino-driven winds release additional $\sim 10^{-4}$ \Msun with $s \sim 40 \; k_{\rm B}$ baryon$^{-1}$
           \cite{dessart09} and a resulting electron fraction set by the neutrino properties, 
           $Y_e^f \simeq [1 + L_{\bar{\nu}_e} \langle \epsilon_{\bar{\nu}_e} \rangle / L_{\nu_e} \langle 
           \epsilon_{\nu_e} \rangle ]^{-1}$ \cite{qian96b} and
\vspace*{-0.3cm}
   \i[iii)] viscously/nuclearly disintegrated accretion disks deliver another  $ \sim 3 \times 10^{-2}$ \Msun
            with $Y_e$ from 0.1 to 0.5 with a peak around 0.3 and entropies from 3 - 20 $k_{\rm B}$ baryon$^{-1}$ 
           \cite{metzger09b}.
\ei
\vspace*{-0.3cm}
The first nucleosynthesis calculations \cite{freiburghaus99b} based on the hydrodynamic ejecta trajectories from neutron 
star mergers \cite{rosswog99} yielded (for an initial $Y_e \sim 0.1$) an excellent agreement with the observed observed 
r-process distribution beyond Ba. Subsequent, 
inhomogeneous chemical evolution studies \cite{argast04}, however, disfavored neutron star mergers
as {\em dominant} r-process sources. This judgment was based on low occurrence rates which
produced an r-process enrichment and a scatter in [r-process/Fe] ratios considered inconsistent with
observations at low metalicities. In the meantime the best estimates for
merger rates have increased by about an order of magnitude \cite{kalogera04a} and it has been realized, 
both from population synthesis, e.g. \cite{belczynski06}, and the observations of a short GRB 
at a redshift of z= 0.923 \cite{graham09}, so just shortly after the time when most stars were being 
assembled in galaxies, that an interesting fraction of DNS systems could possibly merge very early in the cosmic history.\\
% discuss overproduction of r-process
Given the above numbers for released material and merger rates, a natural concern is whether r-process like material 
could be overproduced with respect to the Galactic observations. It first needs to be said that for most of the above mass 
loss mechanisms no thorough nucleosynthesis calculations exist yet and conclusions concerning their nucleosynthesis mainly 
rely on estimates on the final $Y_e$. With respect to a possible overproduction, Metzger et al. \cite{metzger08} 
suspected that this could point to low initial disk masses. These could come about, for example, by DNS mass ratios 
being very narrowly clustered around unity, since such systems produce the least massive disks and the smallest amount 
of dynamic ejecta. For black hole neutron 
star systems, Newtonian and pseudo-relativistic simulations \cite{rosswog04b,rosswog05a}, came to the conclusion 
that for much of the parameter space the formation of massive disks is not possible anyway since the tidal disruption 
of the neutron star occurs close to or even inside of the last stable orbit. This conclusion may need modification 
once fully conclusive GR calculations exist, but so far they do not yet seem to agree on the size of possible disk 
masses \cite{etienne08,etienne09,shibata09}. Given that it is likely only
a small fraction of the NSBH parameter space that is able to produce sizeable accretion disks (the exception may be systems
with low bh masses and large bh spins), combined with the possibly low event rates, NSBH systems may only play a minor role in
the chemical enrichment of galaxies. Another possibility that would avoid a conflict 
with the stellar enrichment history of the universe, is that much of the ejecta do not end up inside of their host 
galaxies, but, being ejected at large velocities predominantly, but not exclusively, in the outskirts of 
galaxies \cite{fryer99a}, would escape to intergalactic space.
%Ways out:
% i) really r-process? --> calculate
% ii) low disk masses --> metzger
% iii) escape velocities ---> end up in intergalactic space

\subsection{GRBs}
\label{sec:GRB}
% * general picture: robust
The potential of compact binary mergers, either in the form of DNS \cite{blinnikov84,paczynski86,goodman86,eichler89} or 
NSBH systems \cite{paczynski91,narayan92}, as central engines of GRBs has been realized more two 
decades ago. Today they are still arguably the most likely central engine for the short-hard variety of Gamma-ray bursts 
(sGRBs), long-soft bursts (lGRBs) seem to be robustly linked to the death of very massive 
stars. There is by now strong evidence that we are seeing the manifestations of (at least) two different types of central
engines, although the classification scheme into long ($>$ 2s) and short ($<$ 2 s) is by no means incontestable: some
bursts belong formally in one class, but possess properties usually attributed to the other \cite{fynbo06,galyam06,thoene08}.\\
An observational breakthrough for sGRBs came with the first detection of afterglows in summer 2005 by the SWIFT
mission \cite{gehrels05,bloom06}, and like in the case of type Ia supernovae, host galaxies provided decisive hints on the
nature of the progenitors. Already the first few detections showed that sGRBs can occur in elliptical galaxies without
star formation and therefore supported the idea that sGRBs are, in contrast to lGRBs, not directly related to the deaths of
massive stars, but instead can be triggered (at least in some cases) by  an old stellar population.
%, consistent with sGRBs arising from compact binary mergers. 
As of now, there are indications of sGRBs occurring in at least three types of galaxies: i) low redshift ($z<0.5$), high-mass
($L \sim L^\ast$), early-type galaxies and galaxy clusters, ii) low redshift, sub-$L^\ast$, late-type galaxies and iii) faint,
star-forming galaxies at $z>1$, not too different from the hosts of lGRBs \cite{gehrels09}. Overall, sGRBs seem to occur in
galaxies that possess higher metalicities than those of lGRBs \cite{berger09}. Recent HST observations show that sGRBs 
have a median projected offset of about 5 kpc
from the center of their host galaxies \cite{fong10}. Again, this would be consistent with an DNS origin, where at least in a substantial
fraction of cases, the binary having been ``kicked'' in the last supernova explosion, can travel substantial distances before 
merging.  Moreover, no supernova explosion seems to go along with sGRBs \cite{castro_tirado05,fox05,hjorth05,bloom06}, although in
exceptional cases, the kick could accelerate the inspiral and cause a high-energy transient, possibly a short gamma-ray burst, 
typically within a few days of the supernova \cite{troja10a}.
SWIFT has now detected more than 50 sGRBs, their redshifts lie between 0.2 and 2,
with a mean of 0.4, while $0.009 < z < 8$ for long bursts with a mean of 2.3 \cite{gehrels10}. Typically short bursts show an isotropic energy of
$\sim 10^{51}$ erg, about two orders of magnitude lower than for lGRBs. With the estimated (but uncertain) jet opening half-angles
of $\theta_j \sim 15^\circ$ for short and $\sim 5^\circ$ for long bursts \cite{burrows06,grupe06} this translates into true energies
of $\sim 10^{49} -  10^{50}$ ergs for short and $\sim 10^{50}-10^{51}$ ergs for long bursts \cite{gehrels10}.\\
All these observations are consistent with the basic idea that sGRBs originate (at least predominantly) from compact binary mergers.
There are, however, several open issues in the sGRB compact binary connection. First, despite the tremendous progress on the 
theoretical side during the last decade, see Sec.~\ref{sec:approaches}, the related challenges have so far prohibited models to go 
the full way from the central engine to the production of the detectable emission. For example, the central engine works on scales of 
$\sim 10^7$ cm, while the ``prompt'' $\gamma$-radiation is thought to be produced by internal shocks at a distance 
$R_{\rm is} \sim c \Delta t \Gamma^2 \sim 3 \times 10^{14} \; {\rm cm} \; (\Delta t/s) \; (\Gamma/100)^2$, where $\Delta t$ 
is the variability time scale of the central engine and $\Gamma$ the outflow Lorentz factor. The 
GRB ``afterglow'', in contrast, is produced near the deceleration radius, $R_{\rm dec} \sim 3 \times 10^{16} {\rm cm} \; 
(E/10^{52} \; {\rm erg})^{1/3} (100/\Gamma)^{2/3} (1 {\rm cm}^{-3}/n_{\rm ext})$, where the swept-up matter has substantially 
slowed down the blast wave \cite{piran04}. Moreover, the central engine requires physics (strong gravity, high-density matter 
physics) that is very different from the emission regions (collisionless, relativistic plasma shocks). 
Up to now, there is not even a model that self-consistently shows the production of the required ultra-relativistic outflow.  
This is partially due to the difficulty to numerically resolve the relevant scales, for example the growth length scale of the 
MRI \cite{balbus98} or the density contrasts between the central parts of a merger remnant and its surrounding 
``near-vacuum'' material, and due to the lack of appropriate 3D neutrino-radiation-hydrodynamics codes that include 
apart from cooling also neutrino heating and annihilation processes.\\
Maybe the most difficult challenge for the compact binary merger model is the emission long after the burst: SWIFT has 
observed X-ray flares in both long and short bursts, which in the latter case occur tens or even hundreds of seconds 
after the burst, corresponding to $\sim 10^5$ dynamical time scales of central compact object. A variety of different 
physical interpretations
has been offered, including refreshed shocks \cite{panaitescu06}, magnetic regulation of the accretion flow \cite{proga06}, the
interaction with a stellar companion \cite{macfadyen05,dermer06}, the (at least temporary) survival of a highly magnetized
central neutron star \cite{rosswog03c,dai06,gao06,rosswog07b,rowlinson10} and the fallback from neutron star debris returning
to the central object at late times \cite{faber06b,rosswog07a,lee07}. After the initial dynamical phase of a merger, 
the importance of hydrodynamic forces in the nearly unbound fallback material ceases and the material becomes essentially ballistic until
it returns  towards the central remnant at late times. Since the ballistic phase far from the central remnant dominates 
the return time scale, the latter can be estimated in an analytical way from the final matter configuration of a numerical 
simulation. The resulting time scales are encouragingly long (e.g. Fig. 3 in \cite{rosswog07a}), but the model is too 
simple to predict, say, the detailed structure of flares. If at fallback the circularization radii are large enough, 
such material could form another, larger disk with possibly much longer time scales, $\sim 100$ s \cite{lee07}, and for the right set of 
parameters can possibly produce signatures similar to the observed ones \cite{cannizzo09}. While fallback can likely 
explain some of the flaring activity, it may be overstrained in explaining the most extreme cases such as 
flares $\sim 12$ h after the burst observed in GRB050724 \cite{campana06,grupe06}. Similarly
problematic are those short bursts where the fluence in the ``extended emission'' exceeds the one of the initial short spike.
For example, in GRB 080503  the fluence of the extended emission exceeds the spike fluence by a factor of about
30 \cite{perley09}. It is at least not obvious how this should be accommodated within the standard merger model, where one expects more mass
in an initial accretion disk than in the fallback material. Some of the magnetar-like models would be less strained to explain these 
extreme events, but in the light of the essentially unsolved baryonic pollution problem discussed in Sec.~\ref{sec:nu_wind}, it is not
obvious how such events would produce the ultra-relativistic outflow for the prompt emission in the first place. \\
The problem of late time flares containing a lot of energy may be less problematic for dynamical compact object {\em collisions}
instead of gravitational-wave-driven binary mergers. A recent study \cite{lee10} finds that the collision rates should be
substantially larger than previously estimated and comparable to merger rates. Moreover, the authors find that fallback material
in the form of tidal tails can contain in some cases more mass than the initial accretion disk. A burst produced by the disk
and flares triggered by fallback could in such a scenario possibly explain late-time flares that are more energetic than the initial
burst.\\
These
are interesting open issues that need further thought and work. In this context, it may be worth reiterating that one should
remain open-minded about the central engines and that different progenitors may (at some rate) contribute to the
observed population of sGRBs.
% * theoretical models
%      progress in:
%                        a) dynamic spacetime evolution
%                        b) further physics inclusion
%                             - nuclear EOS
%                             - neutrino physics
%                             - magnetic fields
%  * open issues
%     a) direct link to observations missing
%         none of today's models shows self-consistently relativistic outflow
%         problems: - time scales (ns <--> internal shock)
%                            - length scales (resolve relevant magnetic processes)
%                            - different type of physics required (say, nuclear matter EOS vs. collisionless shocks)
%     b) "baryonic pollution problem" unsolved
%     c) (How) can compact mergers explain "late time activity"
%         - fallback?
%         - second disk (Canizzo) from fallback?
%         - central magnetar activity?
%    --> none of the issues is more severe than with other models
%    --> compact binary mergers still "best buy", but
%          stay open-minded about other possibilities

\begin{figure}[tbp] %  figure placement: here, top, bottom, or page
   \centering
   \includegraphics[width=6cm]{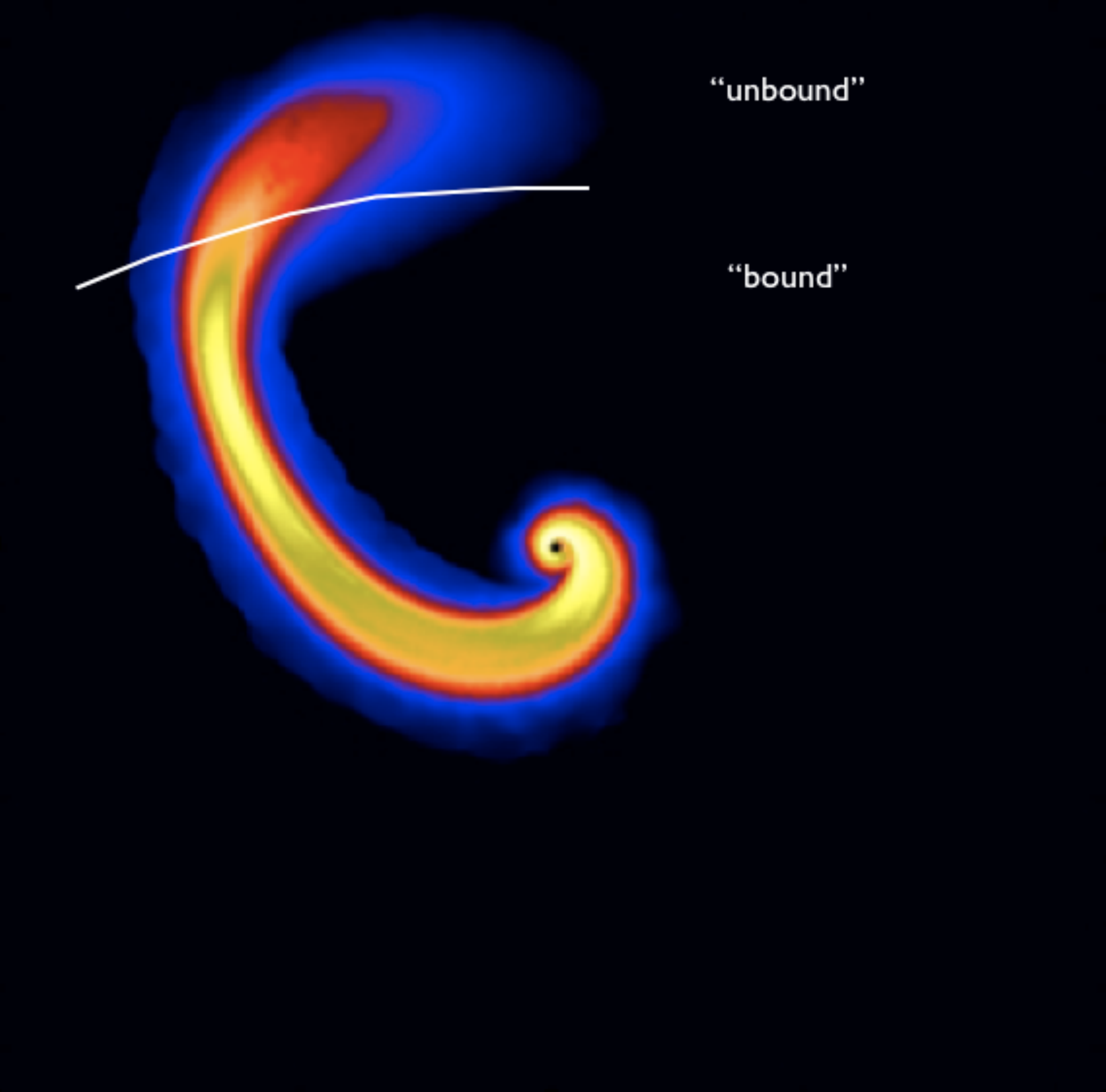} 
   \caption{In the course of a compact binary merger, gravitational torques
            launch 0.02 - 0.08 \Msun into highly eccentric, but still bound
            orbits. This material will fall back on a time
            scale that is much longer than the dynamical time scale of the 
            central remnant. This is illustrated at the example of a neutron
            star black hole merger \cite{rosswog05a}.}
   \label{fig:example}
\end{figure}

\subsection{Pre- and post-merger transients}
\label{sec:transients}
The predicted compact binary merger detection rates for ground-based detector facilities span a broad 
range \cite{kim05,oshaughnessy10} and could be as low as just a few events per year, even for the advanced versions of 
LIGO and VIRGO \cite{smith09}. Therefore, 
additional electromagnetic signals accompanying the gravitational wave emission from compact binaries could 
be crucial to extract as much scientific information as possible from individual detections, e.g. 
\cite{schutz86,bloom09b}.  For example, if position and time could be determined independently, degeneracies 
in the gravitational wave signal could be removed \cite{hughes03,arun09} and the signal-to-noise ratio 
required for a confident detection could be significantly reduced \cite{kochanek93,dalal06}. Moreover, additional
signatures that accompany sGRBs could help to finally unambiguously identify their central engines.\\
Recently, electromagnetic precursor events have been detected for short GRBs \cite{troja10b}. These could,
for example, be the results of the interaction of the neutron star magnetospheres prior to the merger \cite{hansen01}.
This mechanism would require at least one neutron star to be highly magnetized, 
therefore such precursors would not be expected in every merger event.\\
Electromagnetic emission {\em after} such a coalescence is expected from the radioactive nuclei that form via 
decompression of neutron star matter \cite{li98,rosswog99,rosswog05a,kulkarni05,metzger10b}.
GRBs seem to produce highly beamed prompt emission, in contrast to the expected isotropic electromagnetic 
transients that result from radioactive decay in the merger debris. Thus, while the GRB prompt emission can 
only be detected when it points towards Earth, the subsequent, radioactively powered transients should, provided 
they are bright enough, be visible from all directions. Under rare circumstances this may lead to ``orphan transients'' 
when the GRB prompt emission is directed away from us. With the recent and near-future instruments such as the Palomar 
Transient Factory (PTFS \cite{kaiser02}, SkyMapper \cite{keller07}, the VLT Survey 
Telescope (VST) \cite{mancini00}, the Large Synoptic Survey Telescope (LSST) \cite{strauss10} and the Synoptic 
All Sky Infrared Imaging (SASIR) survey \cite{bloom09a}, we are now entering an era of large-scale  
surveys which could revolutionize our understanding of transient events. The detection of a radioactive transient 
coincident with a short GRB could lead to the unambiguous identification of the central engine and provide 
constraints on the r-process production site. 

\section{Summary and future directions}
The past decade has seen a tremendous progress in our ability to reliably model the mergers of compact
binary systems. On the one side, fully relativistic 3D simulations have become possible and are now 
performed regularly. On the other side, many  physical processes have been explored from (exotic)
high-density nuclear physics, over both neutrino cooling and heating to nuclear reactions and magnetic 
fields.\\
Technically, the ``hydrodynamics plus gravity'' part of the models is likely to see a couple of substantial 
improvements. Eulerian approaches are today substantially hampered in their ability to accurately follow lower density
material, to a large extent due to heavy computational burden, so that it is difficult to afford large computational
volumes. Although some refinement techniques are already used \cite{baiotti08,anderson08a}, the field will most likely profit
a lot from the implementation of fully adaptive mesh refinement schemes. Closely related is the treatment  
of ``numerical vacuum'' and its impact on low-density regions. The latter include also the accretion disks
which are thought be crucial in the transformation of gravitational binding energy into observable radiation.
As outlined above, this technical progress will have major implications for both nucleosynthesis and GRB 
questions. Another natural improvement would be the exploration of higher order methods, though there is a 
tradeoff between the order of the method and the affordable resolution.\\
Lagrangian, and in astrophysics this usually means Smoothed Particle Hydrodynamics, methods, are 
currently somewhat lagging behind in their treatment of dynamical space-time evolution with CFA
currently being the most advanced gravity approach \cite{faber06a,oechslin02,bauswein10}. The method
has recently made much progress with the most advanced sets of relativistic equations following directly
from ideal fluid Lagrangians \cite{monaghan01,rosswog09b,rosswog10b,rosswog10a} and 
hybrid approaches that couple SPH with grid-based space-time solvers seem promising for an interesting 
class of problems. Moreover, adaptive Lagrangian-Eulerian (ALE) approaches may be worth the development 
effort.\\
Also on the non-gravity side remains much to do. Problems where the hard to reach numerical resolution
is the major stumbling block need apart from massive parallelism further algorithmic developments and
suitable approximation methods. Today, still many simulations are run with polytropic equations of state
with fixed adiabatic exponents which are insufficient approximations for the required broad spectrum of 
physical conditions in a merger. With resolution becoming better, the simulated times becoming longer and 
low-density phenomena such as winds now becoming feasible, also the presently available physical/nuclear 
equations of state reach their limits, mainly at low densities and temperatures. Closely related is the 
question of nucleosynthesis, another crucial facet in the multi-messenger
picture of compact binary mergers. Despite its importance it is still in its infancy and many 
pressing questions are essentially unexplored. For some related technical problems, say the implementation 
of nuclear reactions into existing hydrodynamics codes, the development effort should be moderate. For others,
say for self-consistent 3D calculations of neutrino-driven winds, a serious amount of work is needed to achieve 
computationally affordable and reasonably accurate results.\\
In anticipation of the first {\em direct} gravitational wave detection, one can overall be optimistic that 
the field will keep its high current momentum, so that hopefully in time for the first gravitational wave
detections reliable multi-physics models will be in place.

%\begin{thebibliography}{99}
%\bibitem{...} 
%\end{thebibliography}

\bibliography{astro_SKR}
\bibliographystyle{elsart-num-sort.bst}

\end{document}